\documentclass[sigconf]{acmart}

\usepackage{color}
\usepackage{float}
\usepackage{subcaption}
\usepackage{xspace} 
\usepackage{multirow}
\usepackage{array}
\usepackage{enumitem}

\definecolor{burntorange}{rgb}{0.8, 0.33, 0.0}
\definecolor{byzantium}{rgb}{0.44, 0.16, 0.39}
\definecolor{byzantine}{rgb}{0.74, 0.2, 0.64}

\AtBeginDocument{%
  \providecommand\BibTeX{{%
    \normalfont B\kern-0.5em{\scshape i\kern-0.25em b}\kern-0.8em\TeX}}}

\copyrightyear{2022} 
\acmYear{2022} 
\setcopyright{rightsretained} 
\acmConference[CHI '22]{CHI Conference on Human Factors in Computing
Systems}{April 29-May 5, 2022}{New Orleans, LA, USA}
\acmBooktitle{CHI Conference on Human Factors in Computing Systems
(CHI '22), April 29-May 5, 2022, New Orleans, LA, USA}
\acmDOI{10.1145/3491102.3501968}
\acmISBN{978-1-4503-9157-3/22/04}

\newcommand{\bolderandunderline}[1]{\textbf{\underline{#1}}}

\newcommand{\systemname}{Crystalline\xspace} 
\newcommand{\systemnameInFull}{\bolderandunderline{C}lipping \bolderandunderline{R}esulting in \bolderandunderline{Y}our \bolderandunderline{S}tructure as \bolderandunderline{T}ables \bolderandunderline{A}nd \bolderandunderline{L}ists
\bolderandunderline{L}inked to
\bolderandunderline{I}mplicit \bolderandunderline{N}otetaking \bolderandunderline{E}asily\xspace}
\newcommand{\stackoverflow}{Stack Overflow\xspace}
\newcommand{\unakite}{Unakite\xspace}
\newcommand{\javascript}{JavaScript\xspace}

\newcommand{\authorplusetal}[1]{#1 et al.\xspace}

\newcommand{\userquote}[1]{``\textit{#1}''}

\begin{document}

\title{\systemname: Lowering the Cost for Developers to Collect and Organize Information for Decision Making}


\author{Michael Xieyang Liu}
\affiliation{%
  \institution{Human-Computer Interaction Institute, Carnegie Mellon University}
  \city{Pittsburgh, PA}
  \country{USA}
  }
\email{xieyangl@cs.cmu.edu}

\author{Aniket Kittur}
\affiliation{%
  \institution{Human-Computer Interaction Institute, Carnegie Mellon University}
  \city{Pittsburgh, PA}
  \country{USA}
  }
\email{nkittur@cs.cmu.edu}

\author{Brad A. Myers}
\affiliation{%
  \institution{Human-Computer Interaction Institute, Carnegie Mellon University}
  \city{Pittsburgh, PA}
  \country{USA}
  }
\email{bam@cs.cmu.edu}


\begin{abstract}
Developers perform online sensemaking on a daily basis, such as researching and choosing libraries and APIs. Prior research has introduced tools that help developers capture information from various sources and organize it into structures useful for subsequent decision-making. However, it remains a laborious process for developers to manually identify and clip content, maintaining its provenance and synthesizing it with other content. In this work, we introduce a new system called \systemname that automatically collects and organizes information into tabular structures as the user searches and browses the web. It leverages natural language processing to automatically group similar criteria together to reduce clutter, and uses passive behavioral signals such as mouse movement and dwell time to infer what information to collect and how to visualize and prioritize it. Our user study suggests that developers are able to create comparison tables about 20\% faster with a 60\% reduction in operational cost without sacrificing the quality of the tables.
\end{abstract}





\maketitle

\section{Introduction}
Developers spend a large portion of their time searching and making sense of the web for solutions to their programming problems \cite{brandt_two_2009,sadowski_how_2015}. In many cases, the answers to such problems are not limited to a single solution, but developers discover that there are multiple legitimate options, and they must identify relevant criteria and constraints based on their unique contexts and carefully consider the trade-offs among those possible options \cite{liu_unakite:_2019,liu_reuse_2021,hsieh_exploratory_2018,patil_comprehensive_2016,gizas_comparative_2012,ratanaworabhan_jsmeter:_2010,peguero_empirical_2018,lawrence_comparing_2017,rutar_comparison_2004,lei_performance_2014}. For example, when converting an old web application to use a modern \javascript front-end framework, React.js \cite{facebook_react_2018} (with its ability to be progressively adopted into existing code bases) may be more suitable when one wants to gradually convert each separate module while minimizing the overall system downtime, whereas a more comprehensive framework such as Angular \cite{google_angular_2019} might be a better choice if one wants to take advantage of various official utility packages like routing \cite{google_angular_nodate}, animation \cite{google_angular_nodate-1} and data validation\break \cite{google_angular_nodate-2}.

There have been many commercial and research tools and systems that try to help people make sense of information about trade-offs to facilitate further decision making, such as by helping with easily capturing snippets of information \cite{bharat_searchpad_2000,schraefel_hunter_2002,noauthor_google_2012,vermette_social_2017,hahn_bento_2018} from web pages or organizing and synthesizing information into useful schema and representations \cite{liu_unakite:_2019,dontcheva_summarizing_2006,ward_orgbox_2021,chang_mesh_2020,kittur_standing_2014,horvath_understanding_2021}. For example, one common practice that people employ is copying pieces of text as well as taking screenshots and putting them in a running Google Doc as they search and browse the web \cite{palani_conotate_2021}. One system that is relevant to the context of programming is \unakite \cite{liu_unakite:_2019}, which enables developers to collect and organize information online into comparison tables with options, criteria, and evidence to help with making decisions (see Figure \ref{fig:original-unakite-figure}).

 \begin{figure*}[t]
 \centering
 	\includegraphics[width=0.82\linewidth]{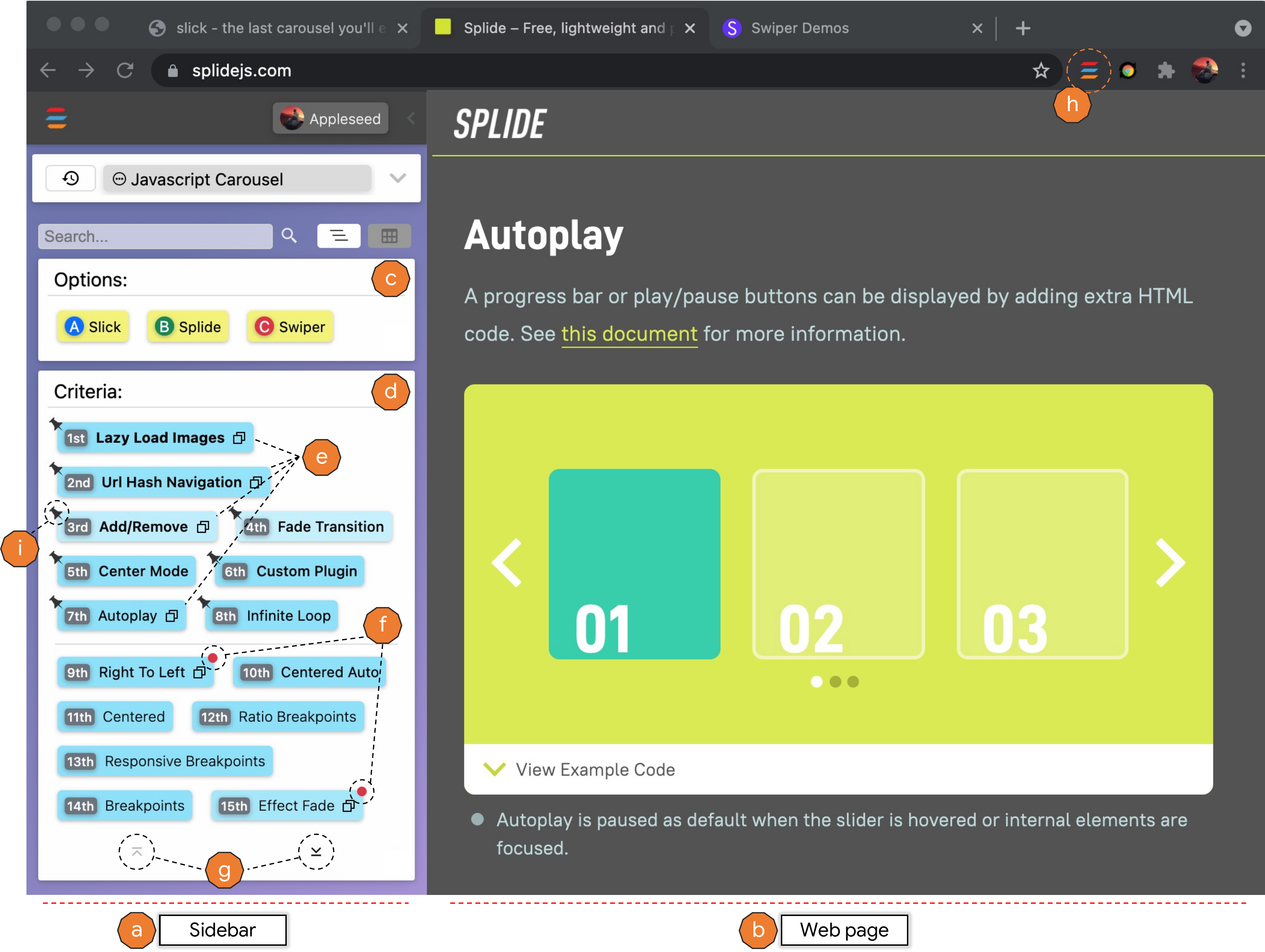}
 	\vspace{-2mm}
 	\caption{\systemname's list view UI (a). As the developer browses a web page (b), \systemname attempts to automatically collect options and criteria from the page, and display them in the options (c) and criteria panes (d) in the sidebar (a). In addition, \systemname leverages natural language processing to automatically group similar criteria together, as shown by the multiple-pages icon (e).  \systemname uses behavioral signals such as mouse movement and dwell time to try to automatically detect the relative importance of the criteria (shown by the display order, with most important at the top). Users can use the ``See more'' and ``See less'' buttons (g) to adjust how many criteria are to be displayed at once. \systemname will remind users of the existence of additional related evidence through a red notification dot at the top right of a criterion (f). The sidebar can be toggled in and out by clicking the browser extension icon (h). Users may pin (i) important criteria to the top of the list.}
 	\Description{\systemname's list view UI (a). As the developer browses a web page (b), \systemname attempts to automatically collect options and criteria from the page, and display them in the options (c) and criteria panes (d) in the sidebar (a). In addition, \systemname leverages natural language processing to automatically group similar criteria together, as shown by the multiple-pages icon (e).  \systemname uses behavioral signals such as mouse movement and dwell time to try to automatically detect the relative importance of the criteria (shown by the display order, with most important at the top). Users can use the ``See more'' and ``See less'' buttons (g) to adjust how many criteria are to be displayed at once. \systemname will remind users of the existence of additional related evidence through a red notification dot at the top right of a criterion (f). The sidebar can be toggled in and out by clicking the browser extension icon (h). Users may pin (i) important criteria to the top of the list.}
 	\label{fig:teaser-image}
 \end{figure*}

However, even with the above tools, it remains a challenging process for developers to \textit{manually} identify and capture the relevant content, maintain its provenance (where it came from), and synthesize it with other content. Prior work suggests that one cause is that people are often uncertain about which information will eventually turn out to be relevant, valuable, and worth capturing, especially at early stages of their learning and exploration when they are overloaded with information \cite{fisher_distributed_2012,bawden_perspectives_1999}. Under these circumstances, people are hesitant to frequently pause and shift their focus from the investigation itself to reasoning about what to capture for later use \cite{chang_supporting_2016,schilit_beyond_1998,kittur_costs_2013,hinckley_informal_2012}, or they could be too engaged in the sensemaking process and forget to collect anything at all. Indeed, research suggests that interactions for gathering information while performing active reading need to be quick and low effort, otherwise people tend not to capture information in the first place \cite{hinckley_informal_2012,marshall_saving_2005,tashman_active_2011,liu_unakite:_2019}. In addition, though existing tools provide users with the flexibility and agency to synthesize the collected information into useful representations, such as comparison tables \cite{liu_unakite:_2019,chang_mesh_2020} or knowledge maps \cite{nguyen_sensemap_2016}, developers still need to perform these organizing operations manually. This is often a laborious process, as developers need to take stock of all the pieces of information, identify connections among them, and directly manipulate the representation to reflect the connections.

Another challenge reported in prior work is that developers' needs for collecting and organizing information are often not discovered until part of the way through an investigation process \cite{liu_unakite:_2019,chang_tabs_2021}. This could be due to several major reasons, including but not limited to: 1) additional external requirements, constraints, or user feedback are discovered or introduced in the middle of a project which significantly complicates the original decision making problem \cite{dyba_empirical_2008,cockburn_agile_2001,dzvonyar_context-aware_2016}; 2) developers discover many more options, criteria, and their trade-offs than they anticipated at the beginning \cite{liu_unakite:_2019}; and/or 3) developers are required to explain or document their decisions and design rationale after the fact for the long-term maintainability and success of a software project \cite{latoza_hard--answer_2010,sillito_questions_2006,latoza_maintaining_2006,lethbridge_how_2003,roehm_how_2012,forward_relevance_2002,de_souza_study_2005}. In these situations, it is hard and involves duplicate work for developers to recall and retrace their steps for reaching their current state of sensemaking (the linear history visualization in almost all current browsers is known to be not particularly effective \cite{won_contextual_2009,kandala_framework_2018,chang_tabs_2021}) and recollect all the relevant evidence again.

In our new work, we explore the idea of having a system dynamically help users keep track of and organize information by leveraging the content they are browsing and the signals from their browsing behavior. Although we focus on the domain of programming due to strongly motivating prior work and ease of prototype development due to regularities of the programming context, our work may also generalize to other sensemaking contexts on the web. We instantiate this idea in a prototype system called \systemname,\footnote{\systemname is named after rocks made up of interlocking crystals. It stands for \systemnameInFull.} which is an extension to the Chrome web browser. \systemname plays the role of a user's copilot and attempts to automatically identify and keep track of the options, criteria, and the corresponding evidence snippets from the web pages that a user has viewed, and organize the snippets into both list and tabular formats. To achieve this, \systemname mines a variety of behavioral signals while a user browses the web, including scrolling patterns and mouse cursor actions, and employs natural language understanding techniques to automatically classify and organize the collected content. The goal is that users can focus more on reading and understanding web content while occasionally guiding the system when it makes mistakes. We conducted a user study to evaluate the usability and effectiveness of \systemname compared to \unakite as a baseline, which found that developers are able to build comparison tables about 20\% faster with a 60\% reduction in operational cost without sacrificing the quality of the tables. In particular, it only requires around 12\% of the total task completion time for participants to use the tool to build and maintain a table, compared to around 30\% in the baseline condition. 

The primary contributions described in this paper include:
\begin{itemize}
    \item evidence that it is possible to automatically identify options, criteria, and relevant evidence from web pages that a user is browsing using a set of natural language understanding heuristics,
    \item a set of implicit behavioral signals that users exhibit when browsing the web which can be used for prioritizing and filtering that collected information,
    \item a prototype system called \systemname that integrates the heuristics and signals to automatically collect and organize viewed information into list and comparison table views for subsequent decision making,
    \item an evaluation that offers empirical insights into the usability, usefulness, and effectiveness of those signals and the system.
\end{itemize}

\section{Related Work}\label{sec:rw}

\subsection{Sensemaking in Software Development}\label{sec:rw-sensemaking-in-software-dev}

Sensemaking is widely considered to be the process of searching, collecting, and organizing information to iteratively develop a mental model that best fits the evidence \cite{pirolli_sensemaking_2005,russell_cost_1993}. As knowledge workers \cite{brandt_two_2009}, many activities that developers perform on a daily basis involve extensive sensemaking, such as designing the overall software architecture \cite{hesse_supporting_2013,maccormack_exploring_2006}, learning and understanding unfamiliar code and concepts \cite{ko_six_2004,deline_easing_2005}, debugging and fixing incorrect software behaviors \cite{ko_exploratory_2006,de_souza_study_2005}, planning and executing code refactorings \cite{murphy-hill_how_2012,fowler_refactoring_2018,ernst_measure_2015}, and evaluating past code and design patterns for future reuse \cite{park_facilitating_2021,liu_reuse_2021}. In this work, we focus on the particular type of sensemaking activity where developers leverage web resources to make a \textit{decision} to solve their programming problem \cite{brandt_two_2009,hsieh_exploratory_2018}. Here, developers not only need to find information pertinent to their problem \cite{brandt_example-centric_2010,ponzanelli_seahawk:_2013,hoffmann_assieme:_2007,stylos_mica:_2006}, which is the first step in such complex sensemaking tasks \cite{russell_cost_1993,white_supporting_2006}, but also collect and synthesize relevant information into structured knowledge so that they can make progress towards fully understanding the decision space \cite{hahn_bento_2018,liu_unakite:_2019,kittur_standing_2014,kittur_costs_2013}. Indeed, our survey \cite{hsieh_exploratory_2018} revealed that over half of the questions asked on \stackoverflow contain answers with multiple options, each option valuable to the programming community due to a unique set of criteria that it fulfills.

Software engineering research has also identified that subsequent developers frequently need help with understanding the rationale of design decisions and code implementations made by previous developers \cite{latoza_hard--answer_2010,sillito_questions_2006, latoza_maintaining_2006}. This can be particularly difficult if the previous developers failed to properly document the rationale \cite{van_de_vanter_documentary_2002}, or the documentation was incomplete or not up-to-date \cite{fluri_analyzing_2009}. 
Granted, the fundamental challenge here is that it is effort- and time-intensive for decision authors to document their rationale (either in situ or after the fact) with little immediate payoff for themselves \cite{gizas_comparative_2012}. Our previous \unakite tool \cite{liu_unakite:_2019} addressed this challenge by encouraging authors to document their decision making processes and results using the tool's lightweight collecting and organizing features.
Building on top of this, \systemname further transforms the previously active capturing and organizing work \cite{bharat_searchpad_2000,schraefel_hunter_2002,liu_unakite:_2019} 
into passive monitoring and error-fixing \cite{li_fmt_2019}, which has been shown to present a much lower entry barrier for people to start contributing \cite{fisher_distributed_2012}.

\subsection{Tools for Collecting and Organizing Information}

To help people more effectively gather and process online information, systems and tools like SenseMaker \cite{baldonado_sensemaker:_1997}, SearchPad \cite{bharat_searchpad_2000}, Hunter Gather \cite{schraefel_hunter_2002}, CoSense \cite{paul_cosense:_2009}, Tabs.do \cite{chang_tabs_2021}, as well as commercial systems like the Evernote clipper \cite{evernote_best_nodate}, enable people to take entire pages or snippets of content from the web, classify them, and later put them together into a document with a coherent narrative for sensemaking, decision making or sharing and collaboration. However, one common characteristic of these tools is that it is mostly the user's responsibility to \textit{manually} complete the information collection, triage, and organization process, while we attempt to do this \textit{automatically} with \systemname as the user searches and browses the web.

Other threads of prior research have explored different ways for machines to help during sensemaking, which inspired and informed our design. For example, systems like Entity Quick Click \cite{bier_entity_2006,ives_interactive_2009,stylos_citrine_2004} employ techniques like named-entity recognition \cite{mansouri_named_2008} to pre-process and highlight semantically meaningful entities in web content, and enable users to collect and annotate relevant information with a single click. Previous work like Thresher \cite{hogue_thresher_2005} and \authorplusetal{Dontcheva}'s personal web summarization tool \cite{dontcheva_summarizing_2006}  let users annotate and curate patterns and templates of information that they would like to collect on a few example web pages, then automatically collect them from future pages. In addition, \authorplusetal{Chang}' Mesh system \cite{chang_mesh_2020} automatically retrieves relevant consumer product facts and reviews from Amazon into a comparison table to enable users to curate and explore nuanced options and criteria. These systems have largely relied on natural language understanding to analyze and transform the web content that users browse and read, while we argue that leveraging the signals from users' natural browsing behavior, such as dwell time and cursor movements, would unlock a new design space for automated machine support during online sensemaking, motivating us to use both NLP heuristics and passive behavioral signals to infer what information to collect and how to visualize and prioritize it in \systemname.

\begin{figure*}[t]
\centering
	\includegraphics[width=1\linewidth]{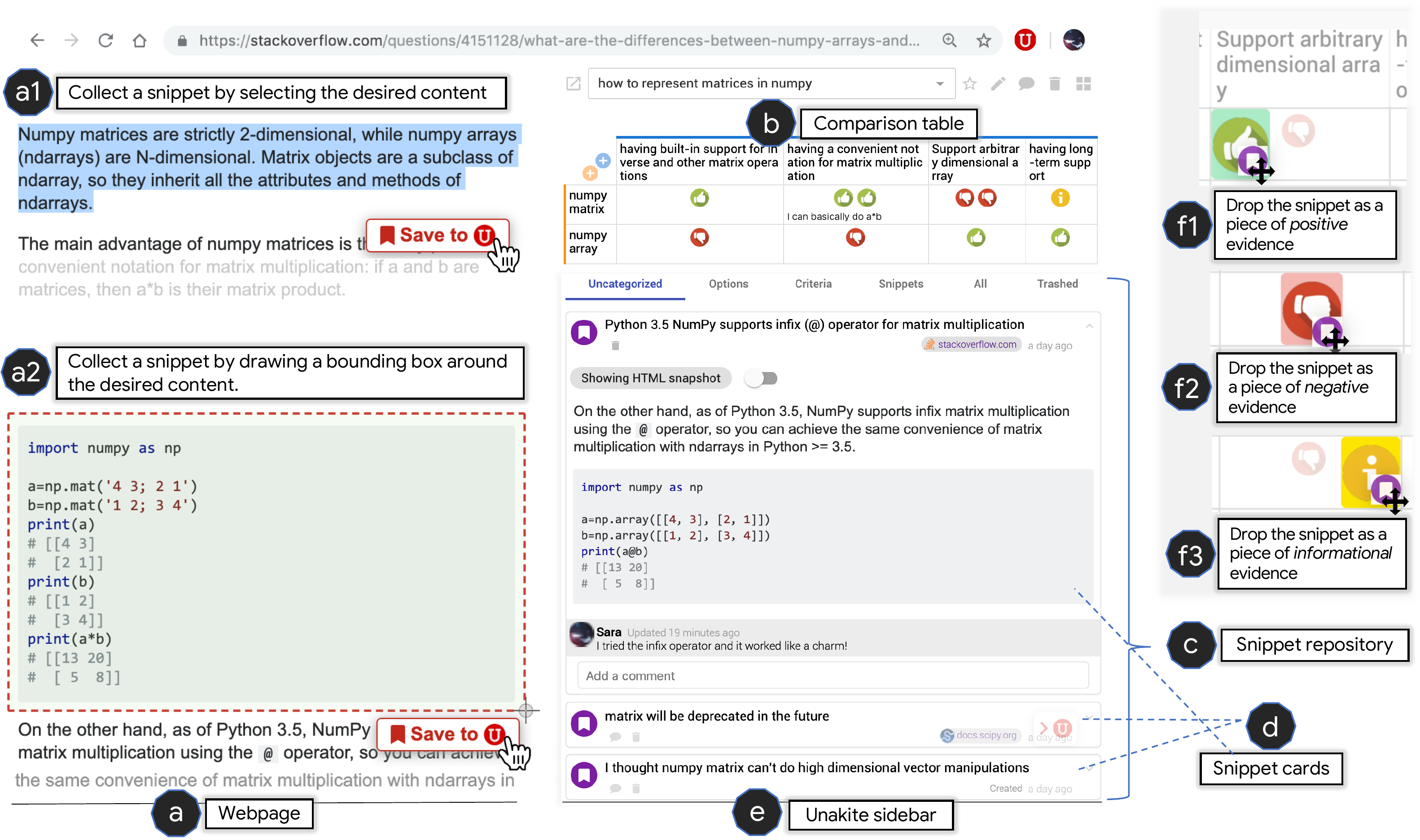}
	\caption{\unakite's user interfaces. With \unakite, a developer collects snippets by selecting the desired content (a1) or by dragging out a bounding box around the desired content (while holding the \texttt{Option} / \texttt{Alt} key) (a2) and clicking the ``Save to U'' button. The collected snippets will show up under the ``Uncategorized'' tab in the snippet repository (c) inside the \unakite sidebar (e). The developer can drag a snippet and drop it in one of the cells in the comparison table (b), and mark whether it is positive (green thumbs-up) or negative (red thumbs-down) or just informational (yellow ``i''). (f1-f3) show the details of the three parts of each cell in the table where the snippet can be dropped. This figure is adapted from \cite{liu_unakite:_2019}. For full details, see \cite{liu_unakite:_2019}.}
	\Description{\unakite's user interfaces. With \unakite, a developer collects snippets by selecting the desired content (a1) or by dragging out a bounding box around the desired content (while holding the \texttt{Option} / \texttt{Alt} key) (a2) and clicking the ``Save to U'' button. The collected snippets will show up under the ``Uncategorized'' tab in the snippet repository (c) inside the \unakite sidebar (e). The developer can drag a snippet and drop it in one of the cells in the comparison table (b), and mark whether it is positive (green thumbs-up) or negative (red thumbs-down) or just informational (yellow ``i''). (f1-f3) show the details of the three parts of each cell in the table where the snippet can be dropped. This figure is adapted from \cite{liu_unakite:_2019}. For full details, see \cite{liu_unakite:_2019}.}
	\label{fig:original-unakite-figure}
\end{figure*}

\subsection{Implicit Behavioral Signals When Using the Web}

Prior research has investigated various implicit behavioral patterns that people exhibit when reading and interacting with content on a digital screen. One thread of research has explored using behaviors such as dwell time, cursor movements, clicks, scrolling patterns, and gaze positions as \textit{implicit signals} to approximate user interest on web pages as well as search result relevance \cite{claypool_implicit_2001,hijikata_implicit_2004,huang_improving_2012,guo_exploring_2008,guo_ready_2010}. For example, \authorplusetal{Claypool} \cite{claypool_implicit_2001} had participants use a custom-built browser to surf the web and concluded that the time spent on a page, the amount of scrolling on a page, and the combination of time and scrolling had a strong correlation with explicit user interest. In addition, Hijikata \cite{hijikata_implicit_2004} 
discovered that actions such as text tracing and link pointing are decent behavioral indicators for perceived interesting segments of web pages. Similarly, in the domain of web searches, \authorplusetal{Buscher} \cite{buscher_what_2009,buscher_eye_2008,buscher_segment-level_2009}, Guo and Agichtein \cite{guo_exploring_2008,guo_ready_2010}, and \authorplusetal{Huang} \cite{huang_improving_2012} demonstrated that eye tracking, as well as interactions like scrolling and cursor hovers, could accurately predict user interests in search results pages.

Building on the empirical understanding laid out by this research, in this work, we explore putting a combination of these implicit behavioral signals into use to approximate user visual attention in a working prototype. We used heuristics and pilot testing to devise mechanisms that translate the raw behavioral signals into numeric scores representing the ``amount of attention'' a user has given to a particular piece of online content. We then use these scores to filter out and rank the content of the evolving comparison table, further reducing the cost for developers to manually manage and prioritize collected information incrementally as they are searching and browsing.


\section{Background and Design Goals}
In this work, we explore automatically keeping track of and organizing relevant information on the web about trade-offs for developers as they are making decisions. To ground our research, we build on the ``Option-Criterion-Evidence'' framework introduced in our \unakite system \cite{liu_unakite:_2019}. We first briefly explain this framework as well as the \unakite system to provide necessary background for this research. Then we discuss the design goals for the new \systemname system.

\subsection{The \unakite System}

\unakite was designed to address both the need of developers to synthesize online information about trade-offs when making\break programming decisions as well as the need of subsequent developers to be able to understand the rationale behind those decisions \cite{liu_unakite:_2019}. As a Chrome extension, \unakite enables developers to manually collect any content from any web pages as \textit{snippets} (pieces of information, Figure \ref{fig:original-unakite-figure}-d) into the snippet repository (a holding tank of information snippets, Figure \ref{fig:original-unakite-figure}-c) by selecting (Figure \ref{fig:original-unakite-figure}-a1) or dragging out a bounding box to enclose the desired content with the mouse cursor (Figure \ref{fig:original-unakite-figure}-a2). To organize the collected content, developers can use drag-and-drop to move the collected snippets from the repository into a comparison table (Figure \ref{fig:original-unakite-figure}-b) \textit{options} (as row headers, e.g., a solution to solve a problem), \textit{criteria} (as column headers, e.g., a standard by which options are judged), and \textit{evidence} (``thumbs-up'' or positive, ``thumbs-down'' or negative, and ``informational'' (``i'') ratings that spread across the rest of the table cells) that illustrates the trade-offs among various options on those criteria. Developers can also rank the options and criteria in the table to reflect their unique order of preferences. The resulting comparison table is automatically saved and can be used by subsequent developers to understand the context of the previous decision space: what options and alternatives were explored, what criteria needed to be met, what trade-offs were discovered, and what was considered the most important and why.

Although \unakite has been shown to incur less operational overhead when it comes to collecting and organizing information in situ compared to common baseline methods like using Google Docs \cite{liu_unakite:_2019}, developers still need to manually collect and structure each piece of content, which can be a costly process \cite{hinckley_informal_2012,marshall_saving_2005,tashman_active_2011,kittur_costs_2013,kittur_standing_2014}. In addition, it forces developers to start using the tool from the outset to be able to capture the whole exploration, but, for cases in which the needs for collecting and organizing information are not discovered until partway through an investigation process (which can be quite common in agile style software development \cite{liu_unakite:_2019,cockburn_agile_2001,dyba_empirical_2008,dzvonyar_context-aware_2016} that is widely adopted across the software development industry), developers would have to retrace their exploration paths from the beginning and re-collect and organize the content, wasting time and causing duplicate work.

\subsection{Design Goals}

In order to address the above limitations of \unakite as well as other similar sensemaking tools \cite{baldonado_sensemaker:_1997,bharat_searchpad_2000,paul_cosense:_2009,chang_tabs_2021}, we formulated the following design goals:

\begin{itemize}
    \item \textbf{Minimize the cost to collect information.} The system should attempt to automatically collect information in the background without the user's specific attention or direction. This will help users focus on the main task of reading and comprehending the content.
    
    \item \textbf{Actively filter, organize, and prioritize information.} The system should actively filter, organize, and prioritize the collected information that gets presented to the user and help the user avoid information overload.
    
    
    \item \textbf{Reduce the cost of incorrect automation support.} In cases where machine support is incorrect or undesirable, the system should allow users to easily recover from those mistakes \cite{horvitz_principles_1999,amershi_guidelines_2019}.
\end{itemize}

\begin{figure*}[t]
\centering
	\includegraphics[width=\linewidth]{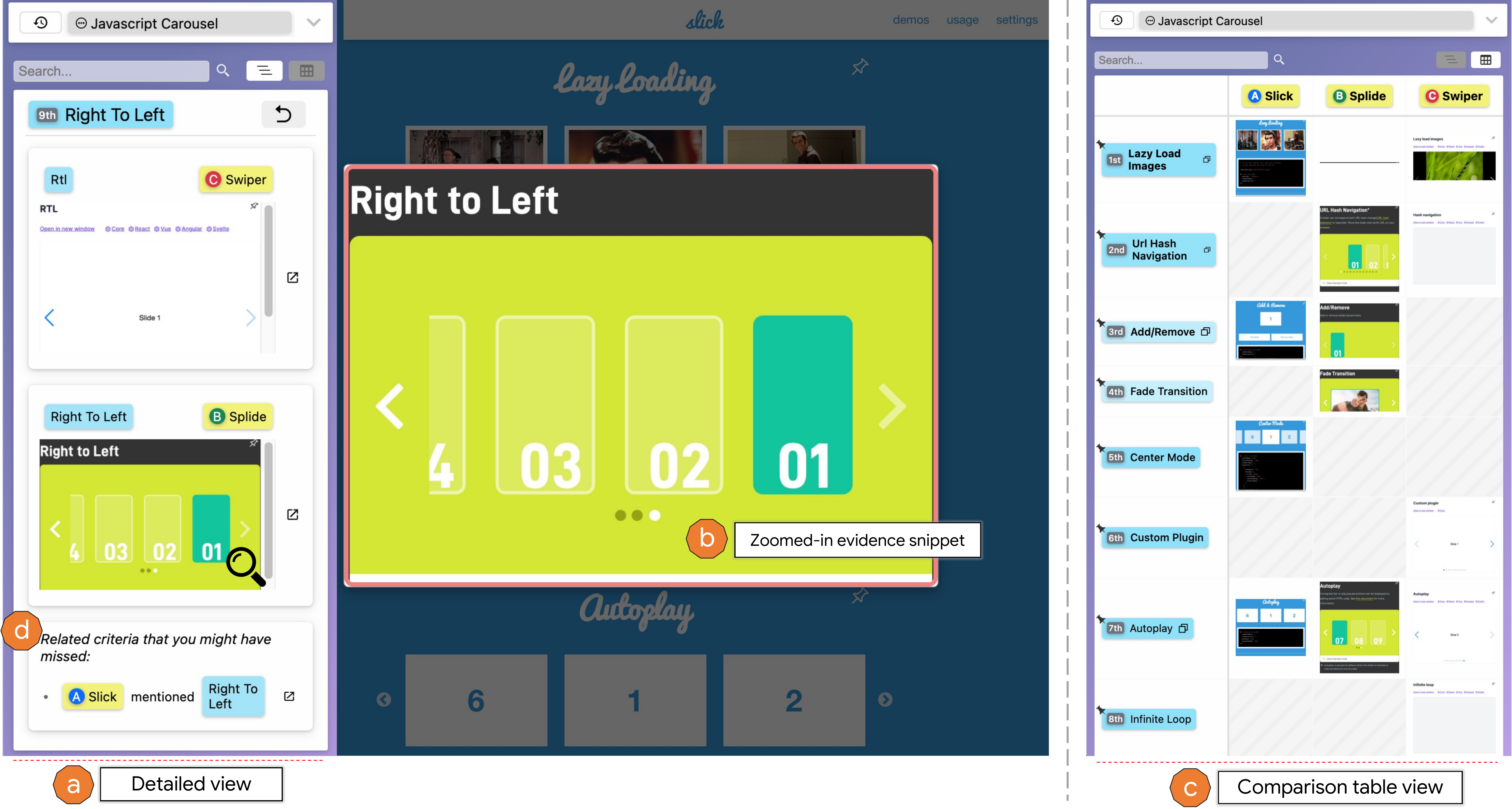}
	\vspace{-5mm}
	\caption{Additional \systemname's user interfaces. Clicking on one of the criterion in the criteria pane (Figure \ref{fig:teaser-image}d) will enter a detailed view for that criterion (a), listing out all the collected evidence snippets organized by options. Users can zoom in on an evidence snippet (b) by moving the mouse cursor over it in the detailed view until the cursor becomes a magnifying glass. Crystalline will actively look for and remind users of evidence for the same or similar criteria from pages that users have visited but have not yet paid attention to (d). Finally, similar to \unakite \cite{liu_unakite:_2019}, \systemname offers a comparison table view (c) that summarizes the decision making space and the trade-offs among various options in detail.}
	\Description{Additional \systemname's user interfaces. Clicking on one of the criterion in the criteria pane (Figure \ref{fig:teaser-image}d) will enter a detailed view for that criterion (a), listing out all the collected evidence snippets organized by options. Users can zoom in on an evidence snippet (b) by moving the mouse cursor over it in the detailed view until the cursor becomes a magnifying glass. Crystalline will actively look for and remind users of evidence for the same or similar criteria from pages that users have visited but have not yet paid attention to (d). Finally, similar to \unakite \cite{liu_unakite:_2019}, \systemname offers a comparison table view (c) that summarizes the decision making space and the trade-offs among various options in detail.}
	\label{fig:detail-and-table-view}
\end{figure*}

\section{\systemname}

\subsection{System Overview}
Guided by prior work and our design goals, we designed and implemented \systemname, a Chrome extension prototype to help developers automatically collect and organize information relevant to their decision making problems.

Users mainly interact with \systemname through a \textbf{sidebar} (Figure \ref{fig:teaser-image}a) that is injected directly into every web page. As a developer opens and reads web pages, the sidebar will be updated with the automatically collected options (Figure \ref{fig:teaser-image}c) and criteria (Figure \ref{fig:teaser-image}d) in the \textit{list view} (Figure \ref{fig:teaser-image}c \& d). The list view serves as a concise and glanceable outline that reflects one's exploration progress --- what options one has encountered and what criteria one has looked into. Clicking on one of the criteria will enter a detailed view for that criterion (Figure \ref{fig:detail-and-table-view}a), listing out all the collected evidence snippets organized by options; similarly, clicking on an option will enter the \textit{detailed view} for that option, which lists all the related criteria and the corresponding evidence associated with that option. Details on how we currently implemented the automatic collection and organization features are discussed in section \ref{sec:detailed-design}.

In addition, developers can also switch to the \textit{comparison table view} (Figure \ref{fig:detail-and-table-view}c) that summarizes the decision making space and the trade-offs among various options in detail. The order in which a criterion gets presented both in the list and the comparison table view are based on the \textit{estimated importance} of the item to the user, which we approximate by the \textit{amount of attention} a user has given to it. This, in turn, is derived from the user's implicit behavioral signals, which we will discuss in detail in section \ref{sec:organizing-and-prioritizing-information}. To examine a particular piece of evidence in the detailed view or a comparison table cell, users can hover on it to zoom in (Figure \ref{fig:detail-and-table-view}b), or click on it to \textit{teleport} to the original web page and scroll position from where it was previously collected.

Similar to previous systems \cite{liu_unakite:_2019,rachatasumrit_forsense_2021,horvath_understanding_2021}, the sidebar can be toggled in and out like a drawer by clicking the extension icon (Figure \ref{fig:teaser-image}h) or using a keyboard shortcut. Developers can passively monitor the sidebar as they are searching and browsing to make sure the system performs correctly, and quickly correct or dismiss the mistakes that the system makes. In addition, developers are free to hide the sidebar to have an unobstructed view of the web page, knowing that all the features for automatic information collection and organization are still running in the background even if the sidebar is in the hidden state.

\subsection{Detailed Design}\label{sec:detailed-design}
We now discuss how the different features in \systemname are designed and implemented, and how they support our design goals.

\subsubsection{Collecting information about options and criteria}

In \systemname, we explore having the system \textit{automatically} collect relevant information in the background without the user having to explicitly perform the action of collecting information. This has the benefit of minimizing the distraction and cost of keeping track of information as an extra step in addition to thinking about the content on a web page, which, in turn, maximizes a user's attention to reading and understanding the content itself.

Specifically, \systemname collects information about options, criteria, and their associated evidence snippets as discussed previously, which was reported by prior work as the key aspects developers look for when solving decision making problems \cite{hsieh_exploratory_2018,latoza_hard--answer_2010,liu_unakite:_2019}. Currently, to automatically recognize the \textit{options}, \systemname employs the following techniques: (1) it looks for the word or phrase between any instances of ``vs.'' (or other variants like ``v.s.'', ``versus'', etc.) in web page titles and opening paragraphs and adds them as potential options. For example, the Medium.com article titled ``Tensorflow vs Keras vs Pytorch: Which Framework is the Best?''\footnote{\url{https://medium.com/@AtlasSystems/tensorflow-vs-keras-vs-pytorch-which-framework-is-the-best-f92f95e11502}} would yield ``Tensorflow'', ``Keras'', and ``Pytorch'' as three potential options; (2) it first runs noun phrase and entity extractions using the Google Cloud Natural Language API \cite{google_cloud_2021} on the web page title, section headers as well as the column and row headers of any HTML tables, then checks if the identified entities are mentioned in the titles of other visited pages. In addition, it also checks if the identified entities would frequently come up in each other's Google autocomplete results (the Google ``vs'' technique is described in \cite{foster_google_2020,liu_reuse_2021}, which issues queries in the form of ``\texttt{[option\_name] vs}'' to the Google Autocomplete API to get a list of autocomplete results that can be interpreted as potential alternatives to ``\texttt{[option\_name]}''. An earlier version of this technique was launched as an experimental feature named Google Sets \cite{chitu_google_2011,tong_united_2008}). Furthermore, it checks if the identified entities are mentioned repeatedly across the main content of the current web page.
All potential options will go through a final deduplication process to produce the final list of options presented in the \textit{options pane} (Figure \ref{fig:teaser-image}c) in the sidebar. We chose and tuned these heuristics based on our internal usage and pilot testing results. In the future, more advanced NLP techniques could be used to augment the current set of heuristics. 

\systemname uses a similar set of heuristics to identify \textit{criteria} from the web pages, with an emphasis on examining section headers and table headers (and entities extracted from them) rather than website titles. In this work and in the context of programming, we focus on using such heuristics to identify the criteria directly mentioned in the content, such as extracting ``learning curve'' from ``React is widely considered to have quite a steep learning curve.'' We leave the extraction of latent criteria for future work, which are more commonly seen in domains other than programming, such as extracting ``price'' from ``I bought this mp3 player for almost nothing'' \cite{poria_rule-based_2014}. 

\begin{figure}[t]
\centering
	\includegraphics[width=0.75\linewidth]{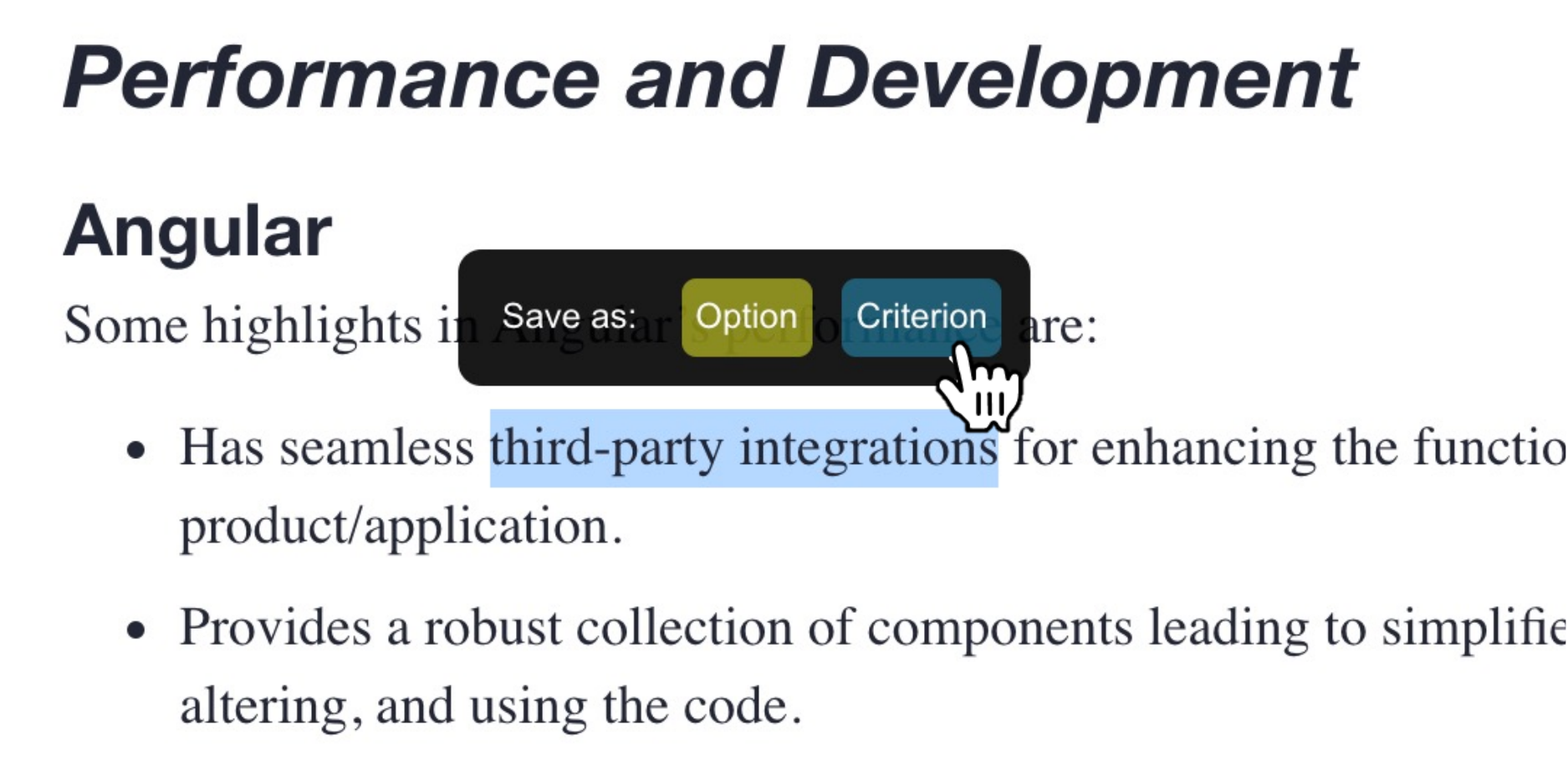}
	\caption{Using the selection popup menu to manually collect options and criteria.}
	\Description{Using the selection popup menu to manually collect options and criteria.}
	\label{fig:selection-popup}
\vspace{-4mm}
\end{figure}

Further, users can always edit the options and criteria names, delete unwanted options or criteria, or manually select and collect any text as either an option or a criterion using the popup menu (Figure \ref{fig:selection-popup}) as a backup.

\def\arraystretch{1}
\begin{table*}[t]
\centering
\resizebox{1\textwidth}{!}{%
\begin{tabular}{
>{\raggedright}p{25mm}|
>{\raggedright}p{36mm}|
>{\raggedright}p{55mm}|
>{\raggedright}p{24mm}|
p{52mm}
}
\toprule
\textbf{Implicit Behavioral Signal} & \textbf{Selected References in Prior Research} & \textbf{Descriptions} & \textbf{Strength of indication of user attention} &\textbf{Score Function $W$} \\ 
\midrule

\textbf{Copying content} & Developers frequently copy sample code from the web to use in their own code \cite{brandt_example-centric_2010,head_interactive_2018,hartmann_hypersource:_2011} &
Triggers when the user copies some text from a content block $b$. This typically happens when a developer copies sample code from web pages to try out in their own code. & 
Strongest &
40 for each triggering \\\midrule

\textbf{Text highlighting} & People tend to highlight text while reading to help focus their attention \cite{roy_note_2021} &
Triggers each time when some text in a content block $b$ gets selected. Triggerings where the selected text is shorter than 5 characters are disqualified. & 
Strong &
20 for each triggering
  \\\midrule

\textbf{Clicking} & Clicking on content, such as widgets and links, is considered to be a decent behavioral indicator for perceived interesting elements on web pages \cite{hijikata_implicit_2004} &
Triggers when the user clicks on a content block $b$. This accounts for situations where the developer interacts with content on a page, such as live demo widgets. Clicks that are part of text highlighting are excluded. &
Strong &
20 for each triggering  \\\midrule

\textbf{Cursor hovering} & People tend to use the cursor to guide their attention while reading web pages \cite{hijikata_implicit_2004,huang_improving_2012,chen_what_2001,guo_towards_2010,rodden_eye-mouse_2008}. &
Triggers each time when the mouse cursor hovers over a content block $b$ for at least 2 seconds. This accounts for situations where the developer naturally moves the mouse cursor onto the content that is currently being read to guide his or her attention \cite{rodden_exploring_2007,rodden_eye-mouse_2008,huang_no_2012,chen_what_2001}.
However, a cursor hover triggering will be disqualified when the system detects an extended period of idling (2 minutes) without any user actions. & 
Weak &
$0.5t$, where $t$ is the duration (measured in seconds) of the cursor's stay within the bounds of content block $b$. The maximum score is 10. In our pilot testing, users rarely spend more than 10 seconds reading a text block. \\\midrule

\textbf{Content dwelling} & The longer some content stays visible, the more likely that the user is interested in it \cite{claypool_implicit_2001,huang_improving_2012}. &
Triggers each time when a content block $b$ gets scrolled into and stays in the visible view port for at least 2 seconds. This indicates that the developer has at least paid attention to $b$. However, a dwell triggering during idling is disqualified. & 
Weak &
$0.2t$, where $t$ is the duration (measured in seconds) of content block $b$'s stay in the visible browser viewport. The maximum score is 4. In our pilot testing, users rarely stay at one location for more than 10 seconds. \\

\bottomrule
\end{tabular}%
}
\vspace{1mm}
\caption{Implicit behavioral signals used in \systemname to track user attention. Column 1 lists the implicit signals; column 2 provides evidence from selected prior research on the efficacy of the signals; column 3 describes how the signals are used in \systemname; column 4 indicates the relative strength of a signal in terms of predicting user attention; column 5 details the scoring function used to translate signal triggerings into numeric scores based on the relative signal strengths. The scoring functions were empirically determined through iterative pilot testing.}
\label{tab:implicit-signals}
\vspace{-5mm}
\end{table*}

\subsubsection{Organizing and prioritizing information}\label{sec:organizing-and-prioritizing-information}
Not all options or criteria are equally useful to a particular developer. Prior work has suggested that a programming decision usually comes down to how well each option matches the developer's goals and criteria that he or she deemed important \cite{patil_comprehensive_2016,gizas_comparative_2012,ratanaworabhan_jsmeter:_2010,peguero_empirical_2018,lawrence_comparing_2017,rutar_comparison_2004,lei_performance_2014,liu_reuse_2021}. In this work, we explore using the amount of attention that one pays to a particular criterion to approximate its perceived value or importance. To operationalize this, for each web page that a developer visits, \systemname processes all the content blocks (HTML block-level elements, such as <p>, <li>, <pre>, and <div>, etc.) to detect what options and criteria are associated with each block. Specifically, it prioritizes verbatim mentioning of options and criteria within a block, then possible options and criteria identified from section headers above the block, then web page titles. If no options are detected, the page title is used as a placeholder. 

Next, \systemname tracks each triggering of five implicit behavioral signals (\textit{copying content}, \textit{text highlighting}, \textit{clicking}, \textit{cursor hovering}, and \textit{content dwelling}) listed in Table \ref{tab:implicit-signals} on any content block and translates it into a numeric score (using column 5). The final attention score $A_c$ representing the amount of attention that a user pays to a particular criterion $c$ is then calculated using equation (\ref{formula:attention-score}):

\begin{equation}\label{formula:attention-score}
    A_c = \sum_{t \in T}I(t,c)\times W(t)
\end{equation}
where $T$ is the set of all implicit signal triggerings; $t$ is a particular triggering; $I(t,c)$ returns $1$ if $t$ was triggered on a content block that is associated with the criterion $c$, and returns $0$ otherwise; and $W(t)$ is the corresponding scoring function found in the last column in Table \ref{tab:implicit-signals}. The scoring functions were empirically determined through iterative pilot testing. 

To accommodate various behavioral patterns exhibited by different users, we iteratively recruited four batches of participants with diverse backgrounds and job responsibilities both within our lab and externally. We followed a diary study approach \cite{rieman_diary_1993} by monitoring their online searching and browsing behavior related to programming through a custom chrome extension that logs triggerings of the above behavior signals and ranks the importance of the associated content blocks accordingly (the initial score functions were determined through our heuristics). At the end of each sensemaking episode, we prompted them to review how well the system did in inferring what they thought was important, and tuned the score function heuristics accordingly (favoring recall over precision). We leave more advanced and adaptive scoring models for future work to investigate. 

By default, the system shows the top 15 criteria ranked by decreasing attention scores in both the list and the table view. Users can use the ``See More'' and ``See Less'' buttons to adjust how many criteria that they would like to see at the same time (Figure \ref{fig:teaser-image}g). As the user browses more content and spreads his or her attention on different content blocks, the order of these criteria changes accordingly in real-time, which provides the user with an ambient awareness of what the system thinks are important. To provide users with the flexibility to override the system's ranking, they can right-click on a criterion and use the ``pin this criterion'' feature to pin it at the top (Figure \ref{fig:teaser-image}i). They can additionally specify their own order of preferences by dragging and dropping to reorder the criteria in the table view, which will automatically pin a criterion if it is not already pinned. Each time an implicit behavioral signal triggering is detected, \systemname also collects the target content block as an evidence snippet, which is presented with its original styling \cite{liu_unakite:_2019} in the detail views and the comparison table view as mentioned above.

\subsubsection{Managing connections and relationships.}

One way for \systemname to actively manage the relationships among the collected information is to automatically merge similar criteria together into \textit{criteria groups} (indicated by a ``multiple items'' icon at the end, see Figure \ref{fig:teaser-image}e). To achieve this, we leverage recent advances in transformer machine learning models such as Universal Sentence Encoder \cite{cer_universal_2018} and BERT \cite{devlin_bert_2019} that can encode textual content into semantically meaningful vector representations called embeddings \cite{goldberg_word2vec_2014}, i.e., two or more semantically close pieces of content will also be close in the embedding vector space (measured by a distance metric, e.g., the cosine similarity distance between vectors \cite{singhal_modern_2001}). \systemname computes an embedding for every criterion as the average of its own embedding and its corresponding evidence snippet, and automatically merges criteria that are within a specified semantic distance threshold to each other into a group. For example, as shown in Figure \ref{fig:detail-and-table-view}a, the system automatically merges ``Right to Left'' (taken from the option ``Splide'') and ``RTL'' (taken from the option ``Swiper'') together since they are semantically similar. The distance threshold was determined empirically through iterative pilot testing. This has the benefit of reducing clutter while helping users make connections among the information that they have seen, which is reported by prior work as one of the difficult steps during sensemaking and schematization \cite{fisher_distributed_2012,russell_cost_1993,pirolli_sensemaking_2005}. In case the system fails to automatically group similar criteria together, users can use drag and drop to manually make the grouping. Similarly, users can easily split a criteria group by right-clicking on the group and hitting the ``split this criteria group'' menu item.

In situations where a user reads and investigates some criterion at one location, \systemname will also actively look for evidence for the same or similar criteria from other pages that the user has visited (including the current page) but has not (yet) paid attention to according to the implicit signals. \systemname will remind the user of the existence of this additional evidence through a red notification dot at the top right of a criterion (Figure \ref{fig:teaser-image}f) as well as in the detailed views (Figure \ref{fig:detail-and-table-view}d). This then serves as an additional way for the system to help users uncover and manage unseen relationships among the information space, as well as a springboard for users to jump directly to the ``overlooked'' information for further investigation.

\subsection{Implementation Notes}
The \systemname Chrome browser extension is implemented in HTML, \javascript, and CSS, using the React \javascript library \cite{facebook_react_2018}. It also uses Google's Firebase for database synchronization and persistence, back-end functions, and user authentication. 

To produce the content embeddings, we used \textit{bert-as-a-service} \cite{devlin_bert_2019} and the \texttt{uncased\_L-12\_H-768\_A-12} pre-trained BERT model to implement a REST API that the extension can query on-demand. 
The embedding calculations are known to incur significant computational costs and delays. Therefore, to ensure a smooth user experience, they are better suited to run on a remote server with the necessary resources rather than locally in an end-user's browser. 

Unlike other systems \cite{pinterest_pinterest_nodate,evernote_best_nodate} that help users find more information from new sources, \systemname only collects information from the web pages that a user has explicitly visited. This is an intentional design choice we make in the current implementation: the major role of \systemname is to remove the burden for users to actively keep track of relevant information that they have personally seen and investigated so that it is easier for them to revisit and recall. We leave the design space of automating the discovery of new relevant information for future research to explore.

\section{Evaluation}
We conducted an initial lab study to evaluate the usability of the \systemname system in helping developers collect and organize information.

\subsection{Participants}

We recruited 12 participants (7 male, 5 female) aged 22-35 ($\mu$ = 27.6, $\sigma$ = 3.7) years old through emails and social media. The participants were required to be 18 or older, fluent in English, and experienced in programming. Participants had on average 6.9 years of programming experience, with half of them currently working or having worked as a professional developer and the rest having programming experience in universities.

\begin{table*}[t]
\centering
\resizebox{1\textwidth}{!}{%
\begin{tabular}{
l| 
>{\raggedright}p{22mm}|
>{\raggedright}p{16mm}|
>{\raggedright}p{14mm}|
>{\raggedright}p{28mm}|
>{\raggedright}p{17mm}|
>{\raggedright}p{20mm}|
>{\raggedright}p{16mm}|
>{\raggedright}p{20mm}|
l
}
\toprule
& 
Manually select information and capture &
Rename an option / criteria &
Delete an option / criteria & 
Manually put information snippets into the table & 
Remove a snippet from the table &
Merge criteria into groups &
Split criteria groups &
Pin or reorder criteria &
\textbf{Overall}
\\
\midrule
    
Task A &
27.0 (6.42) & 
1.67 (1.97) &
0.67 (1.03) &
16.5 (5.43) &
0.50 (0.84) &
N/A &
N/A &
6.00 (2.19) &
52.3 (13.7)
\\
    
Task B &
26.2 (5.56) & 
1.83 (1.60) &
1.50 (1.38) &
14.5 (5.28) &
0.33 (0.82) &
N/A &
N/A &
6.00 (1.79) &
50.3 (14.3)
\\\midrule

\textbf{Average} &
26.6 (5.74) & 
1.75 (1.71) &
1.08 (1.24) &
15.5 (5.21) &
0.42 (0.79) &
N/A &
N/A &
6.00 (1.91) &
51.3 (13.4)
\\
\midrule
\multicolumn{10}{@{}c@{}}{\textbf{(a) \unakite} condition}\\
\end{tabular}%
}
\resizebox{1\textwidth}{!}{%
\begin{tabular}{
l| 
>{\raggedright}p{22mm}|
>{\raggedright}p{16mm}|
>{\raggedright}p{14mm}|
>{\raggedright}p{28mm}|
>{\raggedright}p{17mm}|
>{\raggedright}p{20mm}|
>{\raggedright}p{16mm}|
>{\raggedright}p{20mm}|
l
}
\toprule
& 
Manually select information and capture &
Rename an option / criteria &
Delete an option / criteria & 
Manually put information snippets into the table & 
Remove a snippet from the table &
Merge criteria into groups &
Split criteria groups &
Pin or reorder criteria &
\textbf{Overall}
\\
\midrule
    
Task A &
0.83 (0.75) & 
2.17 (1.17) &
0.50 (0.84) &
0.17 (0.41) &
0.33 (0.52) &
2.33 (0.82) &
0.83 (0.75) &
5.33 (1.97) &
12.5 (3.02)
\\
    
Task B &
1.00 (1.26) & 
1.67 (0.82) &
0.50 (0.55) &
0.33 (0.52) &
0.33 (0.52) &
1.83 (0.75) &
0.67 (0.82) &
5.50 (2.74) &
11.8 (3.31)
\\\midrule

\textbf{Average} &
0.92 (1.00) & 
1.92 (1.00) &
0.50 (0.67) &
0.25 (0.45) &
0.33 (0.49) &
2.08 (0.79) &
0.75 (0.75) &
5.42 (2.27) &
12.2 (3.04)
\\
\midrule
\multicolumn{10}{@{}c@{}}{\textbf{(b) \systemname} condition}\\
\end{tabular}%
}
\caption{Statistics for the average number of interactions performed by users to perform the tasks in the user study. Standard deviations are included in the parentheses.}
\label{tab:num-interaction-table}
\vspace{-5mm}
\end{table*}

\subsection{Procedure}

The study was a within-subjects design, where participants were presented with two tasks and were asked to complete one of them using \unakite (baseline condition) and the other using \systemname (experimental condition), in a counterbalanced order. For each task, participants were presented a programming decision-making problem, a set of four web pages, some necessary background of the problem, and a list of three options available to solve the problem that they were required to investigate. The provided web pages were either documentation pages of specific options or comprehensive review articles reviewing several options together. Participants were instructed to read through the provided web pages, and use either \unakite or \systemname to collect and organize information into a comparison table containing all the given options and at least 8 different criteria in the order of their perceived importance. We imposed a 20-minute limit per task to keep participants from getting caught up in one of the tasks. However, they were instructed to inform the researcher when they have collected 8 criteria as well as the associated evidence. If they wished to continue beyond this checkpoint, they were allowed to, until they felt like they could make no further progress. Specifically, the two tasks were to use the corresponding system in each condition to build a comparison table of:

\begin{itemize}
\item \textit{(A)} Choosing a \javascript carousel library to build a photo sharing web application. The available options were: Splide.js\footnote{\url{https://splidejs.com/}}, Slick\footnote{\url{https://kenwheeler.github.io/slick/}}, and Swiper\footnote{\url{https://swiperjs.com/}}.
\item \textit{(B)} Choosing a front-end framework to implement a basic personal portfolio website. The available options were: React.js\footnote{\url{https://reactjs.org/}}, Angular\footnote{\url{https://angular.io/}}, and Vue.js\footnote{\url{https://vuejs.org/}}.
\end{itemize}

We chose \unakite over other commercially available tools such as Google Docs as the baseline condition because: 1) it can be easily used to capture richer contexts such as formatted text (example code), images, and links; 2) similar to \systemname, it also provides a sidebar that allows participants to view and organize the collected information directly rather than switching context over to another browser tab or application to paste in and structure information; and 3) \unakite was shown to be easy to learn and use in prior research and incurs significantly less overhead cost than using Google Docs \cite{liu_unakite:_2019}.

In addition, rather than letting participants search for their own pages to research, we provided them with the predefined set of pages to ensure a fair comparison of the results, and since helping to find relevant web pages is not a goal of \systemname. Requiring participants to only read the predefined pages (each contains on average 7 screenfuls of content) also helps ensure that the two tasks are of roughly equal difficulty in terms of reading and cognitive processing effort. Furthermore, to ensure realism and participant engagement, the tasks were selected based on actual questions asked and discussed on programming forums and websites. We specifically simplified the requirements and background of task B to match that of task A, since otherwise, choosing a \javascript framework (e.g., to build interactive industry-level web applications) would arguably be more substantial and involve deeper and much more careful comparisons and team discussions that are beyond the scope of this lab study. In fact, as shown in section \ref{sec:qual-results} there was no significant difference by task.

Each study session started by obtaining consent and having participants fill out a demographic survey. Participants were then given a 10-minute tutorial showcasing the various features of \unakite and \systemname and a 10-minute practice session on both systems before starting. At the end of the study, the researcher conducted a survey and an interview eliciting subjective feedback on the \unakite and \systemname experience. Each study session took approximately 60 minutes, using a designated MacBook Pro computer with Chrome, \unakite and \systemname installed. All sessions were carried out in person, with participants and the researcher appropriately masked following COVID-19 mitigation protocols. All participants were compensated \$15 for their time. The study was approved by our institution's IRB office.

\def\arraystretch{1}
\begin{table*}[t]
\centering
\resizebox{1\textwidth}{!}{%
\begin{tabular}{
>{\raggedright}p{128mm}|
p{30mm}|
p{28mm}
}
\toprule
\textbf{Question Statements} &
\textbf{\systemname condition} &
\textbf{\unakite condition} 
\\\midrule

I would consider my interactions with the tool to be understandable and clear. &
6.17 (0.39) & 
6.08 (0.67)
\\\midrule

I would consider it easy for me to learn how to use this tool. &
6.08 (0.79) & 
6.00 (1.04)
\\\midrule

I enjoyed the features provided by the tool. &
6.25 (0.45) & 
6.17 (0.58)
\\\midrule

Using this tool would make solving programming problems at my work more efficient and effective. &
6.08 (0.29)$^*$ & 
5.75 (0.45)$^*$
\\\midrule

If possible, I would recommend the tool to my friends and colleagues doing programming work. &
6.17 (0.58)$^*$ & 
5.58 (0.51)$^*$
\\\bottomrule
\end{tabular}%
}
\vspace{1mm}
\caption{Statistics of scores in the post-tasks survey. Participants were asked to rate their agreement with statements related to their experience interacting with \systemname and \unakite on a 7-point Likert scale from ``Strongly Disagree'' (a score of 1) to ``Strongly Agree'' (a score of 7). Statistics in column 2 and 3 are presented in the form of mean (standard deviation). Statistically significant differences ($p < 0.05$) through paired t-tests are marked with an $^*$.}
\label{tab:survey-scores}
\vspace{-5mm}
\end{table*}

\section{Results}

\subsection{Quantitative Results}\label{sec:qual-results}

All participants were able to complete all of the tasks in both conditions, and nobody went over the pre-imposed time limit. Figure \ref{fig:teaser-image}, together with Figure \ref{fig:detail-and-table-view}, shows an example table built by one of the participants in the study for task A.

To examine how \systemname performs compared to the baseline \unakite condition, we measured the time it took for participants to finish each task. 
A two-way repeated measures ANOVA was conducted to examine the within-subject effects of condition (\systemname vs. \unakite) and task (A vs. B) on task completion time. There was a statistically significant effect of condition ($F(1,20) = 8.06$, $p = 0.01$) such that participants completed tasks significantly faster (21.6\% faster) with \systemname (Mean = 611.8 seconds, SD = 144.6 seconds) than in the \unakite condition (Mean = 780.3 seconds, SD = 137.6 seconds). There was no significant effect of task ($F(1,20) = 0.11$, $p = 0.74$), indicating the two tasks were indeed of roughly equal difficulty. These results suggest \systemname helped participants build up comparison tables faster overall, even the majority of their time was necessarily spent reading through the material in both conditions.

To account for this reading time, we also compared the \textit{overhead cost} \cite{liu_unakite:_2019} of using both tools to collect and organize information. For the \systemname condition, we calculated the overhead cost as the portion of the time participants spent on directly interacting with \systemname (scrolling through the list and table view to examine the evidence collected so far, splitting and merging criteria, pinning important criteria, manually collecting information, etc.) out of the total time they used for a task (vs. reading and comprehending the web pages). Similarly, in the \unakite condition, the overhead cost was calculated as the percent of time participants spent on directly using \unakite features (selecting and collecting information snippets, drag and dropping snippets into the comparison table, etc.), in the same way as was done to compare \unakite to Google Docs \cite{liu_unakite:_2019}.

A two-way repeated measures ANOVA was conducted to examine the within-subject effects of condition (\systemname vs. \unakite) and task (A vs. B) on overhead cost. There was a statistically significant effect of condition ($F(1,20) = 77.5$, $p < 0.001$) such that the overhead cost was significantly lower (almost 60\% lower) in the \systemname condition (Mean = 11.6\%, SD = 0.04) than in the \unakite condition (Mean = 28.4\%, SD = 0.07). Again, there was no significant effect of task ($F(1,20) = 0.53$, $p = 0.48$)). Thus, using \systemname resulted in reduced overhead costs of collecting and organizing information.

To gain deeper insights into \textit{why} the overhead cost was significantly lower in the \systemname condition, we tallied the number of interactions performed in each task while collecting and organizing information to build the comparison tables (Table \ref{tab:num-interaction-table}). Here, we notice that the majority of interactions in the \unakite condition are to manually collect information snippets (on average 26.6 times) and place them into the comparison table (on average 15.5 times). In contrast, in the \systemname condition, the majority of interactions are to merge criteria into groups (on average 2.08 times) and pin or reorder the criteria in the table (on average 5.42 times). This suggests that, to some extent, \systemname has transformed the previously active capturing and organizing work into passive monitoring and error-fixing, which explains the lower overhead cost.

In the survey, participants reported (in 7-point Likert scales) that they thought the interactions with \systemname were understandable and clear (Mean = 6.17, SD = 0.39), \systemname was easy to learn (Mean = 6.08, SD = 0.79), and they enjoyed \systemname's features (Mean = 6.25, SD = 0.45). In addition, compared to \unakite (Mean = 5.75, SD = 0.45), they thought using \systemname (Mean = 6.08, SD = 0.29) would help them solve programming problems more efficiently and effectively, and would recommend \systemname (Mean = 6.17, SD = 0.58) over \unakite (Mean = 5.58, SD = 0.51) to friends and colleagues doing programming work, both differences were statistically significant under paired t-tests. Details of the survey questions and scores are presented in Table \ref{tab:survey-scores}.

\subsection{Qualitative Observations}

\subsubsection{Usability and usage patterns}
Overall, participants appreciated the increased efficiency afforded by various \systemname features. Many (9/12) mentioned that the perceived workload to collect and organize what they have investigated was minimal, saying that \userquote{I feel like I got a table for free} (P3), \userquote{the fact that I can see what I've paid a lot of attention to automatically bubbles up to the top is quite magical} (P9), and \userquote{It feels as if I was sitting in the passenger seat and not having to do all the steering and maneuvering} (P7). Some (3/12) participants also reported having taken advantage of the overlooked information reminder feature (Figure \ref{fig:detail-and-table-view}d) to guide their research. Furthermore, participants reflected that \systemname relieves them of the burden of trying to anticipate the value of a particular piece of information before collecting it since \userquote{the important bits will eventually be at or near the top, hopefully} (P12), and they could \userquote{focus on reading the page itself and not context switch to bookkeeping mode again and again} (P5).

However, some did voice concerns about the system's ability at the beginning of the tasks, arguing that they were \userquote{skeptical if it will actually collect the right things} (P1), and reported that they would \userquote{skim through the list view and the table view quite frequently at the beginning} (P7). However, as they progressed through the tasks, their confidence in \systemname increased, and they only occasionally checked the sidebar. We observed that three of the 12 participants ended up not examining and editing the system's output until they felt like they had finished reading and processing all the given pages, and they made minimal edits to the results.

\subsubsection{Working with machine suggestions}
Participants generally thought that the benefits of automating the collection and organization process outweighed the costs of dealing with occasional unhelpful machine suggestions, such as incorrectly merging criteria together or prioritizing unimportant criteria at the top of the list. For example, P7 reflected, \userquote{it feels like a mind reader. I know it's not perfect, but I also don't expect it to be, and would actually prefer occasionally peeking into what it's been doing and fixing whatever that's not correct than grabbing everything by myself all the time.} 

Some did raise concerns about the ordering of criteria getting changed too frequently (\userquote{they [the criteria] were jumping around}, P7) at the beginning. This is likely due to the fact that users were skimming through a web page without paying particular attention to anything at the beginning, causing their attention scores to be relatively indistinguishable. For future iterations of the system, we could experiment with less frequent UI update intervals under these circumstances so it would cause less distraction.

\subsection{Evaluation Discussion}

Similar to what was reported in prior work \cite{rachatasumrit_forsense_2021}, since our participants were not explicitly told how the system worked to automatically collect and rank information, they had to form their own mental models and hypotheses about how the system works and how they could affect it with their behavior. For example, P8 noticed that \userquote{it looks like if I spend a little bit more time on a particular place on a page, the corresponding criterion would get picked up and bumped up quickly; and if I click on that part a bunch of times, which happens to be what I typically would do when I try to focus my attention on something now that I'm thinking about it, it's [the corresponding criterion] going to go up even faster.} This suggests that our implicit signals were working, and further, that with experience users might adapt to \textit{explicitly} steer the system towards their goal of collecting and prioritizing information, resulting in, to some extent, a mixed-initiative collection approach that still would require much less effort than the baseline methods. Future research could explore the costs and benefits of a wide variety of interactions and signals that lie on the spectrum between implicit behavioral signals to full manual direct manipulations, and any differences caused by directly instructing users about the implicit signals being used.

Though the current version of \systemname mainly focuses on reducing the cost for developers to collect and organize information, which was exactly what we tested in the lab study, we were also interested in making sure that the \textit{quality} of the comparison tables built using \systemname does not degrade as seen in other automation scenarios \cite{grace_viewpoint_2018,shneiderman_human-centered_2020}. Since there is not a gold standard comparison table, we evaluated the correctness of \systemname's automatic approaches by how much editing participants had to do in order to fix \systemname's mistakes and make sure that all the content in the table was eventually filled out and ranked correctly according to their understanding as per the study protocol. As shown in Table \ref{tab:num-interaction-table}(b), participants only had to perform on average 12.2 edits to the automatically generated comparison tables, compared to the 51.3 actions that they had to manually perform in the baseline \unakite condition (the difference is statistically significant, $p < 0.01$). Among these, edits that are related to collecting information, such as manually selecting information and capture (0.92 times), renaming (1.92 times), and deleting information (0.50 times) were minimal, suggesting that our combination of NLP and behavioral signal heuristics was working effectively to collect information that the users thought was important. However, participants pinned or reordered the criteria that were automatically ranked by \systemname on average 5.42 times (SD = 2.27 times). One possible explanation is that the universal scoring functions (in Table \ref{tab:implicit-signals}) did not necessarily apply to every single participant, suggesting the need for a more sophisticated and personalized scoring mechanism in future iterations of \systemname and systems that leverage signals from users' natural browsing behavior.

In addition, we asked and coded their opinions about \textit{using} these tables as if they were the subsequent developers trying to \textit{understand} the design rationale. In general, participants were excited about using comparison tables automatically built by \systemname. For example, P10 highlighted scenarios where \systemname would be useful for his own purposes, saying that \userquote{it's sort of like a never-erased whiteboard that would most likely help me remember what I looked at three months ago.} In addition, some reflected that compared to having no clue of why a decision was made in a particular way in the first place, they would appreciate at least having access to a \systemname table even if it was not actively monitored and maintained during the initial developer's sensemaking process. For example, P4 said: \userquote{I think being able to read something like this [\systemname table] is going to make a big difference when you're banging your head against the wall trying to understand why this particularly old API was chosen, I mean, especially when the guy who wrote the code was long gone, I could at least `read a transcription of his mind' in some sense.} Here, we see preliminary evidence that our approach of automatically collecting and organizing information on behalf of developers is useful and valuable. We leave the formal evaluation of the quality of fully automatically built comparison tables with possibly more advanced versions of \systemname for future work.

\section{Limitations}
Currently, \systemname works best on a limited set of web pages in the programming domain, including documentation pages that are dedicated to a particular library or a set of APIs, as well as review articles or question answering pages that discuss and compare several options together. We chose to optimize for these types of web pages in the current prototype as they are reported in prior work \cite{hsieh_exploratory_2018,liu_unakite:_2019} as well as our formative discussions with developers as some of the most frequently consulted programming resources when it comes to making decisions. However, the performance reported on the web pages used in the study is not necessarily representative of how \systemname would operate even on web pages of these types for users in general. In addition, \systemname currently relies heavily on the overall structure of the web pages being standard, meaning that a page uses HTML tags appropriately according to their semantics (e.g., enclosing headers and list items in \texttt{<h>} and \texttt{<li>} tags rather than wrapping everything with \texttt{<div>} tags) and that there is a strong semantic coherence between a section header and its corresponding content. Though this is sufficient to demonstrate the idea of automatic collecting and organization and the benefits they offer, future research is needed to make \systemname-style tools work on a more diverse set of web pages, as well as how to be clear upfront about its limitations in parsing web pages that do not follow appropriate web standards.

Furthermore, our lab study has several limitations. Given the short amount of training and practice time participants had, some might not have been able to fully grasp the various features of \systemname, or they might have been confused about what \unakite (the baseline system) has to offer. The study tasks might not be what participants typically encounter in their daily work, depending on whether they are in a position to make decisions, and thus they may not be equipped with the necessary motivation or context that they would otherwise have in real life. We mitigate these risks in the study setup by: 1) having participants perform a practice task for each condition simulating what they would have to do in the real tasks; 2) choosing the study tasks based on actual questions that are discussed by developers on \stackoverflow and other popular programming community forums; and 3) providing participants with sufficient background information and context to help them get prepared. In fact, 7 out of 12 participants reported that the tasks were indeed similar to what they would deal with in their daily work.  We would like to further address these limitations in the future by having developers use \systemname on their own work and personal projects, which would provide them with sufficient motivation as well as experience with \systemname enriched over time. 

Finally, the overhead cost measurement in the study could be conservative, as we did not account for the time participants spent simply glancing or looking at the sidebars without any explicit interactions with it. However, from our observations during the study, participants rarely spent any extended time doing this. Nevertheless, we would like to take advantage of more advanced tools such as eye tracking \cite{bigham_--fly_2021,rodden_eye-mouse_2008,papoutsaki_searchgazer_2017,papoutsaki_webgazer_2016} in the future to more accurately account for the proportion of time when a participant's gaze is fixated on the user interface of the tools rather than on actual web content.


\section{Future Work}
Through designing and evaluating \systemname, we gained deeper insights into the benefits and trade-offs of automatically collecting and organizing information for developers as they make sense of the web to make programming decisions. This motivates some ideas for future work.

While \systemname's approach provides developers with an inexpensive way of capturing knowledge in the browser, it represents only one piece of a larger puzzle of how to support a developer's everyday work that involves sensemaking and decision making. One dimension to characterize this is that developers also frequently perform activities outside their browsers, such as in IDEs, code editors \cite{sadowski_how_2015}, command-line interfaces \cite{chen_bashon_2020}, literate programming notebooks \cite{kery_story_2018,kery_variolite:_2017}, or threads of discussions during formal or informal meetings \cite{zhang_making_2018}. Further research would be needed to understand how to collect and organize information from these sources as well as how to integrate them together to provide a more comprehensive picture of the decision making context. Another dimension that is relevant is the lifecycle of the knowledge captured via systems like \unakite and \systemname. Early evidence from the user study has suggested there is a benefit of \systemname's organization from the perspective of a subsequent developer who may need to understand a previous developer's decision. Future research could investigate how well developers are able to understand and potentially reuse these automatically assembled knowledge artifacts, possibly without any manual interventions from the initial knowledge authors, which could, in turn, eliminate the starting cost associated with initial knowledge creation \cite{fisher_distributed_2012} and unlock the virtuous cycle of accelerated programming knowledge reuse \cite{fisher_distributed_2012,liu_reuse_2021}.

Though the current set of mechanisms for deriving the importance of criteria from implicit behavioral signals generally works well for the setting of this research, there could be situations where a user's default browsing behaviors and patterns fall outside the limited set of signals and heuristics that \systemname is currently looking for. For example, a user might not have the habit of unconsciously using the cursor as a reading guide or might not interact with the page at all while reading, which would render the tracking of some of the behavioral signals moot. In addition, users could exhibit different or additional behavior patterns when generalized to other tasks domains that involve information-backed decision making, such as comparison shopping, trip-planning, etc. \cite{hahn_bento_2018,chang_mesh_2020}. For example, when interacting with a map view to find the best local dining option, a user may frequently pan around and zoom (in and out) to view different restaurants, and both the duration of stay on a particular restaurant and how many times it is viewed back and forth could be leveraged to approximate the user's interest and investment of effort. One way to address these concerns is to leverage a more diverse set of behavioral signals and potentially signal combinations, such as scrolling, mouse panning, zooming, eye tracking \cite{papoutsaki_webgazer_2016,papoutsaki_searchgazer_2017,fan_eyelid_2020,fan_eyelid_2021}, and facial gestures tracking \cite{kianpisheh_face_2019,sun_teethtap_2021} to collect a more accurate picture of what users are seeing on screen. Another future direction that could be fruitful is to take a machine learning approach instead of the current rule-based approach for approximating content importance using behavioral signals. Specifically, we could leverage recent advances in crowdsourcing and labeling \cite{deng_imagenet_2009,song_popup_2019,chao_learning_2018,chen_improving_2020} to log, annotate, and construct a large-scale data set that maps a variety of behavioral signals to the perceived importance of content blocks that they are triggered on, and train on this data set to obtain scoring functions that would work more widely. Alternatively, an online learning approach could also be promising, where the system continuously learns, adapts, and\break improves from an individual user's behavior over time, as suggested by Horvitz \cite{horvitz_principles_1999}.

Last but not least, automation afforded by systems like \systemname enable people to focus their attention on reading and comprehending the web pages rather than splitting attention with having to collect and organize the information at the same time. However, prior work in learning science, such as \authorplusetal{Bransford} \cite{council_how_2000}, found that people who \textit{personally} performed the actions of collecting, categorizing, and organizing information were more likely to be able to recall it correctly and in detail, and exhibited increased confidence in the final outcome. This raises an interesting tension and trade-off between full-on automation and direct manipulation --- future research would be required to examine the long term effect on people's learning outcome as well as confidence in their decisions using systems like \systemname, and determine the appropriate levels and circumstances when automatic information bookkeeping should be applied.

\section{Conclusion}

This paper explored how automatically collecting and organizing information as developers search and browse the web can better support them in decision making scenarios. Our designs were motivated by the growing complexity of the decisions that developers need to make, and the lack of tooling support to help them efficiently gather and synthesize evidence without causing much interruption to their main focus of reading and understanding content online. We introduced \systemname, a browser extension that instantiates this idea by leveraging natural language processing and users' behavior signals such as mouse movement and dwell time to infer what information to collect and how to organize and prioritize it on behalf of a user. Through a lab study with 12 participants, we found promising evidence that using \systemname as a copilot to collect and organize information is much faster and more efficient, and the resulting knowledge artifacts are potentially useful and valuable for the initial user as well as for subsequent consumption by people who need to understand the original decision-making context.

\begin{acks}
This research was supported in part by NSF grants CCF-1814826 and FW-HTF-RL-1928631, Google, Bosch, the Office of Naval Research, and the CMU Center for Knowledge Acceleration. Any opinions, findings, conclusions, or recommendations expressed in this material are those of the authors and do not necessarily reflect the views of the sponsors. We would like to thank our study participants for their kind participation and our anonymous reviewers for their insightful feedback. We are genuinely grateful to Yongsung Kim, Joseph Chee Chang, and Amber Horvath for their valuable feedback. In addition, we sincerely thank Jinlei Chen, Tianying Chen, Yulan Feng, Nan Gao, Haojian Jin, Toby Jia-Jun Li, Franklin Mingzhe Li, Julia Jiayin Qian, Haitian Sun, Jiachen Wang, Eric Yiyi Wang, Ziyan Wang, Zheng Yao, and Yi Zhou for their constant support, especially during the COVID-19 pandemic.
\end{acks}


\balance
\bibliographystyle{ACM-Reference-Format}
\bibliography{references}


\end{document}